\documentclass{PoS}
\newcommand{\beqa}{\begin{eqnarray}}
\newcommand{\eeqa}{\end{eqnarray}}
\newcommand{\beq}{\begin{equation}}
\newcommand{\eeq}{\end{equation}}
\newcommand{\sfrac}{{{\textstyle\frac12}}}
\newcommand{\sfracf}{{{\textstyle\frac14}}}
\newcommand{\dtb}{$\overline{\hbox{D3}}$\ }
\def\IK{\relax{\rm I\kern-.20em K}}
\def\IM{\relax{\rm I\kern-.15em M}}

\title{Fine-Tuning in Brane-antibrane Inflation}

\ShortTitle{Fine-Tuning in Brane-antibrane Inflation}

\author{\speaker{James M.\ Cline}\\
        McGill Univeristy, Montr\'eal, Qc H3A2T8, Canada\\
        E-mail: \email{jcline@physics.mcgill.ca}}

\abstract{I give a brief overview of brane-antibrane inflation, with
emphasis on the problems of tuning to get a flat potential in the
KKLMMT framework, and recent work on the nature of superpotential
corrections in that model.}

\FullConference{From Strings to LHC\\
		January 2-10 2007\\
		International Centre Dona Paula, Goa India}

\begin{document}

Brane-antibrane inflation is one of the most important ideas for
inflation from string theory.  I have reviewed it previously in 
\cite{JC1}, to which the reader is directed for more complete
references to the literature. 
Here I will recapitulate some of the historical developments that
led to the KKLMMT \cite{KKLMMT} model, then discuss its tuning
problems, and the challenges for finding superpotential corrections
within string theory which have the right properties for producing a
sufficiently flat potential.

\section{Inflation from brane annhilation} 
\begin{figure}[h]
\centerline{
\includegraphics[width=0.7\hsize,angle=0]{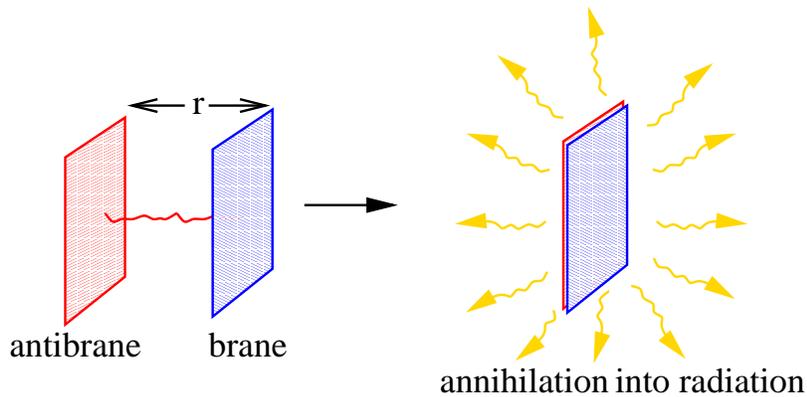}
}
\caption{Brane-antibrane inflation and reheating.}
\label{fig1}
\end{figure}

The interaction energy
between a parallel D3-brane and its corresponding antibrane can give
rise to inflation in the early universe \cite{DT}.  The subsequent
brane-antibrane annihilation  ends inflation and can reheat the
observed universe \cite{BBC}, presumably located on some other brane which may
or may not be coincident  with the inflationary branes.  This is
illustrated in figure \ref{fig1}.  One might wonder whether the
branes being parallel requires an extra fine-tuning beyond those
which will be discussed below.  However for a D3 brane, any nonzero
angle would require the brane to wrap some of the compact dimensions,
similar to a helix on the surface of a drinking straw.  The energy
density of the wrapped brane would be greater than that of a
zero-angle brane due to the greater volume required by wrapping
relative to remaining straight.  Thus the zero-angle configuration is
energetically preferred.  

In this picture, the brane-antibrane separation $r$ plays the role of
the inflaton, and the lightest mode of the stretched string between 
branes becomes tachyonic at a critical separation or order $1/M_s$
(the inverse string mass scale), ending inflation.  In this respect,
brane-antibrane inflation is quite similar to hybrid inflation.

\subsection{Brane-antibrane action}
\label{ss1}

\begin{figure}[h]
\centerline{\includegraphics[width=0.4\hsize,angle=0]{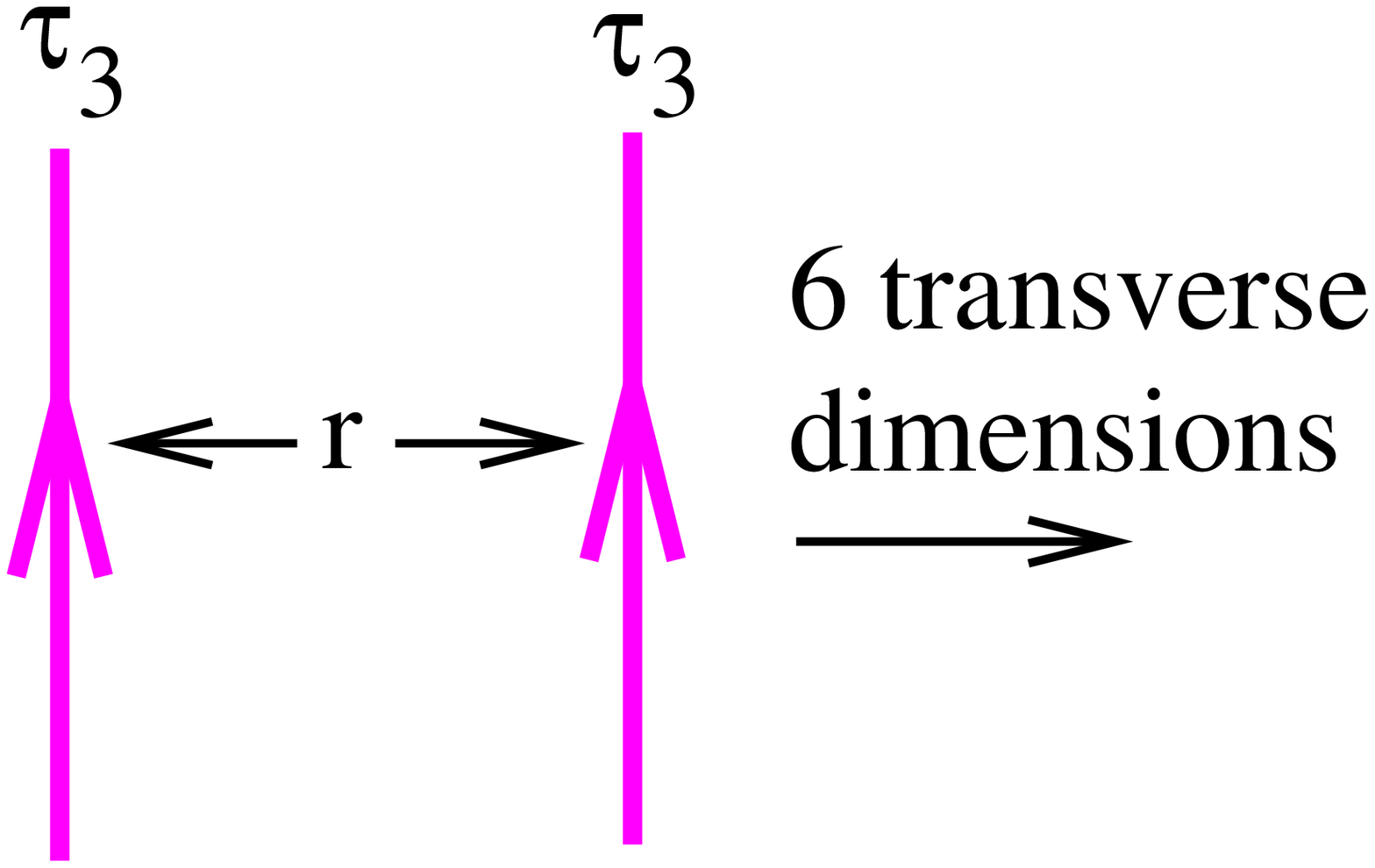}
\hfil
\includegraphics[width=0.4\hsize,angle=0]{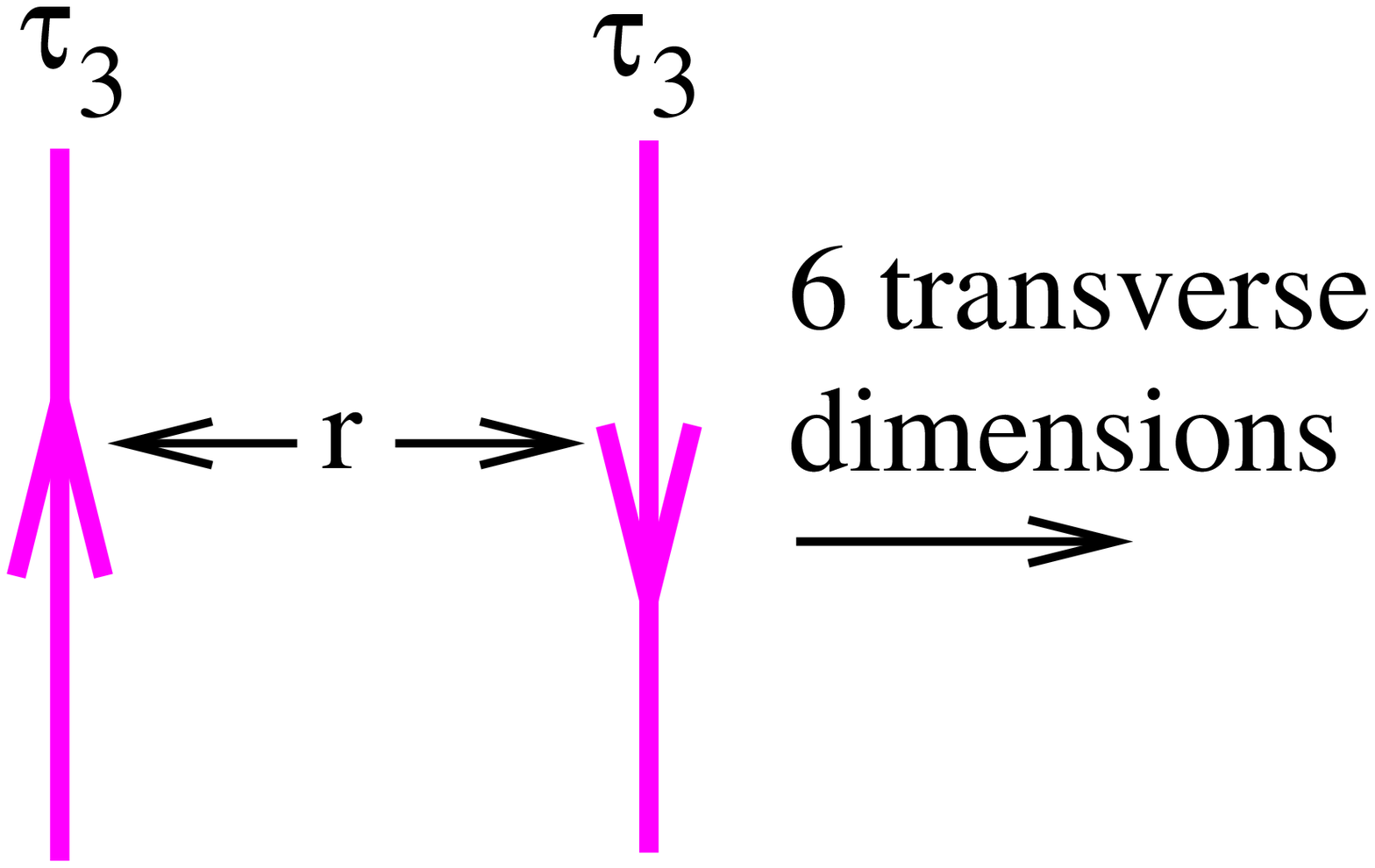}}
\caption{Brane-brane (left) versus brane-antibrane (right)
configuration; orientation hence charge of antibrane is reversed.} 
\label{fig2}
\end{figure}

To understand the inflationary potential, one should first note that
parallel BPS (supersymmetric) D3 branes exert no force on each other.
The two component forces are
\[V_{\rm grav} = - \kappa_{10}^2\,{\tau_3^2\over r^{4}}, \quad \hbox{gravitational
attraction} \]
\[V_{\rm gauge} = + \kappa_{10}^2\,{\tau_3^2\over r^{4}}, \quad \hbox{RR gauge
field repulsion} \]
where $\kappa_{10}^2$ is the 10D gravitational constant, $\tau_3$
the D3-brane tension, and $r$ the separation in the compact
dimensions.  Notice that
due to the BPS condition, these
exactly cancel each other.  On the other hand, for 
antiparallel D3 branes, the orientation and charge of one brane is 
reversed, turning it into an antibrane, as illustrated in 
figure \ref{fig2}.  The gravitational attraction
is no longer canceled by RR-gauge repulsion, resulting in the
attractive total potential
\beq
V_{\rm tot} = 
-2\kappa_{10}^2\,{\tau_3^2\over r^{4}}
\eeq

So far we have treated the brane-antibrane separation $r$ as if it
were a single degree of freedom, but branes are not rigid objects;
they fluctuate
in the transverse directions, so the actual separation is not just a
number, but a field $r(x^\mu)$ which depends on the position $x^\mu$ in the
noncompact directions, as shown in figure 3.

\begin{figure}[h]
\centerline{\includegraphics[width=0.4\hsize,angle=0]{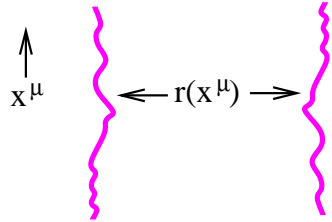}}
\label{fig3}
\caption{The inflaton field $r(x^\mu)$.}
\end{figure}

To find the kinetic term for the inflaton, we start with the
Dirac-Born-Infeld (DBI) action for a single D3 or \dtb brane,
\beq	
	S = -\tau_3\int d^{\,4}x\, \sqrt{-G}
\eeq
where
$G_{\mu\nu}$ is the induced metric on the brane,
\beq
	G_{\mu\nu}= g_{AB} \, {\partial X^A\over \partial x^\mu}\, 
	{\partial X^B\over \partial x^\nu} = \eta^{\mu\nu} + 
	{\partial \phi^I\over \partial x^\mu}\, 
	{\partial \phi^I\over \partial x^\nu}
\eeq
Here $\phi^I$ are the transverse oscillations to the brane, and 
expanding to leading order in them gives  
\beq
	\det G = -1 + \left({\partial \phi\over \partial x}\right)^2+\dots
\eeq
Hence the DBI action takes the approximate form
\beq	S = -\tau_3\int d^{\,4}x\, \left(1 - 
\frac12\left({\partial \phi\over \partial x}\right)^2+\dots
	\right)
\eeq

To find the action for the 
{canonically normalized inflaton}
we let $r^I = \phi^I - \bar\phi^I$, where
$\phi^I,\ \bar\phi^I$ are the respective fluctuations of the brane
and antibrane.  The Lagrangian then splits into an uninteresting
contribution for the center-of-mass, and the relevant one for the
separation,
\beq
	{\cal L} = -\sfrac\tau_3 (\partial r)^2 - V(r)
\eeq
The canonically normalized inflaton is therefore
\beq	\varphi = \sqrt{\tau_3}\, r = \sqrt{\tau_3}\, \left(\sum_I (r^I)^2\right)^{1/2}
\eeq
and its potential is
\beq
	V = 2\left(\tau_3 - {c\over\varphi^4}\right),\qquad c = \kappa_{10}^2\tau_3^4
\eeq
It will be important below that the 10D gravitional coupling is 
\beq	
	\kappa_{10}^2 = M_{10}^{-8} = M_p^{-2}\, L^6
\eeq
in terms of the compactification volume $L^6$.

\subsection{The flatness problem}
To get enough inflation, we need the slow-roll parameters to be small.
One finds that the $\eta$ parameter provides the most stringent
constraint, 
\beq \eta \ \equiv\  M_p^2{V''\over V} \ \sim\ - 
\left({ L}\over r\right)^6 
\label{eta}
\eeq
From this formula, it appears that the only way to make $\eta$ small is to demand that the
brane-antibrane separation satisifes $r \gg L$.  However it is
impossible to separate them by more than the size of the extra
dimensions, so this does not work \cite{cliff}.  In fact the approximation 
(\ref{eta}) is only valid when $r\ll L$; when $r\sim L$
compactification effects become important and the potential no longer
behaves like $1/r^4$ as it does in flat space.  Nevertheless, the
setup is still problematic because of the assumption that the
compactification volume is stabilized.  Realistically $L$ is a modulus
with dynamics that can influence the inflaton.  It is not obvious that
the introduction of a dynamical stabilization mechanism for $L$ will
leave $V(\varphi)$ flat, even if that can be achieved for fixed $L$.  
Thus it is important to have a complete picture in which the dynamics
of compactification is understood.   

\section{Flux Compactifications and the KKLMMT model}

An important step toward more complete and realistic
string-inflationary model building was the realization that
background fluxes can stabilize many of the moduli of string theory.
In particular, 
{ Giddings, Kachru and Polchinski (GKP)
\cite{GKP} showed that fluxes in warped compactifications,
using a Klebanov-Strassler (KS) throat \cite{KS}, 
generically  stabilize the dilaton and complex structure moduli of
 type IIB string theory compactified on a 6D Calabi-Yau manifold.}  The situation
is illustrated in figure 4.  Besides the advantages of moduli
stabilization, this has further appealing features: the throat
generates a hierarchy through warping like in the Randall-Sundrum
(RS) model \cite{RS}; a large hierarchy can be generated from natural
values of the fluxes, which are quantized.

\begin{figure}[h]
\centerline{
\includegraphics[width=0.4\hsize,angle=0]{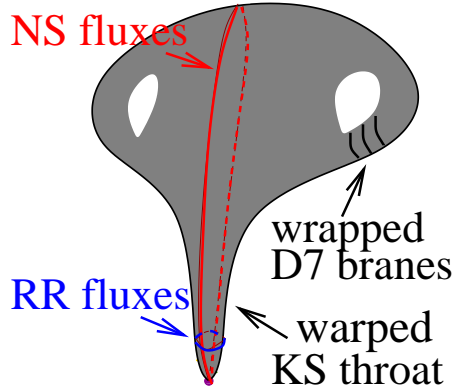}}
\label{fig4}
\caption{Klebanov-Strassler throat attached to a Calabi-Yau manifold, with
fluxes of $H_3$ and $F_3$ wrapping dual 3-cycles.}
\end{figure}

\subsection{Klebanov-Strassler Throat}
\begin{figure}[h]
\centerline{
\includegraphics[width=0.5\hsize,angle=0]{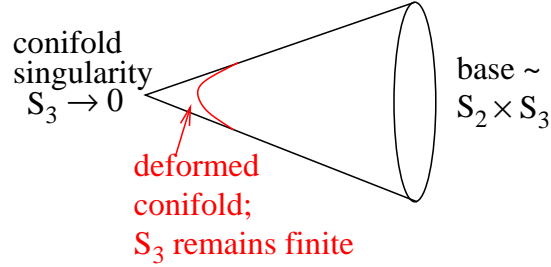}}
\caption{The singular and deformed conifold geometries.}
\label{fig5}
\end{figure}

Let us consider the KS warped throat in more detail.
It can be thought of as a generalization of the RS model
to 10D.  The geometry of the throat is approximately 
AdS$_5\times T_{1,1}$, where $T_{1,1}$ is a compact space described by
five angular coordinates:
\beq ds^2 = a^2(r)(-dt^2 + dx^2) + a^{-2}(r)(dr^2 + r^2 ds_{T_{1,1}}^2)
\eeq
and the warp factor takes the form
\beq	
	a(r) \cong {r\over R},\quad R = \hbox{AdS curvature scale}
\eeq
The throat is a generalization of the singular conifold geometry
pictured in figure \ref{fig5}. It is similar to a cone, but the base 
$T_{1,1}$ has the topology of $S_2\times S_3$ instead of a circle.
At the tip of the cone, where $r=0$, the $S_3$ shrinks to zero size.
One can also consider a deformed conifold in which the manifold closes
off smoothly at some nonzero value $r=r_0$.  These manifolds, which
are complex, can be described in terms of four complex coordinates
$w_i$ restricted by one complex condition,
\beq
	\sum_{i=1}^4 w_i^2 = z
\label{conifold}
\eeq
The case $z=0$ corresponds to the singular conifold, while $z\neq 0$
describes the deformed conifold.  $z$ is a dynamical field, the
complex structure modulus, which is a flat direction in the absence of
fluxes, but which acquires a potential when fluxes are turned on for 
 $H_{(3)}$, the Kalb-Ramond field strength, and for $F_{(3)}$, the
the field strength of the Ramond-Ramond (RR) 2-form
$C_{(2)}$.  The flux quanta are specified by integers $M$ and $K$,
\beq
	\left({M_s\over 2\pi}\right)^2  \int_A F_3 = M,\qquad
\left({M_s\over 2\pi}\right)^2  \int_B H_3 = -K
\eeq
where $A$ and $B$ denote dual 3-cycles of the Calabi-Yau, portrayed
as circles in figure 4.  The stress-energy of the fluxes
 fixes the value of $z$ to be
\beq	
\label{wf}
	z = e^{-2\pi K/g_s M} = a_0^3
\eeq
In language familiar from the RS model, $a_0$ is the 
warp factor at bottom of throat, which plays the role of the infrared brane.

\subsection{Getting Inflation: KKLMMT}

\begin{figure}
\centerline{\includegraphics[width=0.4\hsize,angle=0]{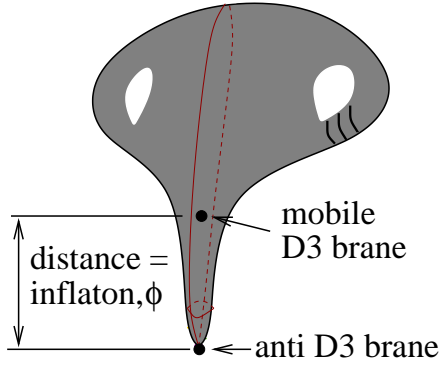}}
\caption{D3 and \dtb in KS throat in KKLMMT setup}
\label{fig6}
\end{figure}

We have now introduced (almost!) all of  the basic ingredients required for
building a semi-rigorous inflationary model from string theory.
KKLMMT \cite{KKLMMT} added a 
D3 and $\overline{\hbox{D3}}$ into the throat, as shown in figure
\ref{fig6}.  In this configuration, the \dtb sinks quickly to bottom
of the throat, while the D3 is almost neutrally buoyant.  This comes
about because of the background fluxes, which induce a RR 5-form
field strength background through its equation of motion,
\beq
	dF_{(5)} \sim H_{({3})}\wedge F_{(3)}
\eeq
The corresponding gauge potential is the 4-form, whose solution is
\beq
	C_{(4)} = a^4(r)
\eeq
The 4-form couples to 
D3 and \dtb  through the Chern-Simons (CS) action, 
{\it i.e.}, the second term in 
\beqa	
\label{sw}
	S &=& -\tau_3\int d^{\,4}x\, \left( a^4(r)
	\sqrt{1 + a^{-4}(r)(\partial\phi^I)^2}
	\mp C_{(4)} \right)\\
	&\cong& \sfrac\tau_3(\partial\phi^I)^2 + \left\{{0,\ \ \ \ \qquad\qquad\qquad\hbox{D3}
	\atop -2\tau_3\, a^4(r) \int d^{4}x, \quad\hbox{\dtb}}\right.
\label{sw2}
\eeqa	
The first term in eq.\ (\ref{sw}) is the DBI action including the warp factor
in the background geometry.  Eq.\ (\ref{sw2}) is leading term in the
slowly-rolling limit.  The constant parts of the DBI and CS terms
cancel for D3 but add for \dtb, explaining why one floats while the
other sinks.  

However, we have ignored the D3-\dtb interaction in the
approximation (\ref{sw2}).  To derive it, one can consider
the action for a static \dtb at position $r = r_0$ in the throat:
\beq
	S = -\tau_3\int d^{\,4}x \sqrt{g_4(r_0)} - 
	\tau_3 \int d^{\,4}x \, \, C_{(4)}(r_0) 
\eeq
If there is no additional brane in the throat, the background fields
have solution $\sqrt{g_4} = C_{(4)} = a^4_0$ and the potential for
the \dtb  is $V= -2a_0^4\tau_3$ as in (\ref{sw2}).  Now imagine adding
a D3 at position $r$; it perturbs the geometry
\beq
   g^{(6)}_{\mu\nu} \to g^{(6)}_{\mu\nu} + \delta g^{(6)}_{\mu\nu}
\eeq
The perturbation satisfies the Poisson equation in the 6 extra 
dimensions,
\beq
	\nabla^2  \delta g^{(6)}_{\mu\nu} = C\, \eta_{\mu\nu}\, 
	\delta^{(6)}(\vec r) \quad \Rightarrow \quad
	\delta g^{(6)}_{\mu\nu}\sim C\, \eta_{\mu\nu}\, (r-r_0)^{-4} 
\eeq 
Substituting the perturbed background $g_4\sim 1/g_6\sim C_{(4)}^2$ 
back into the action (\ref{sw}), one obtains the potential
\beq 
	 V = {2 a_0^4\tau_3 \over 1 + a_0^4(r-r_0)^{-4}}
\label{Veq}
\eeq 
If eq.\ (\ref{Veq}) was the final result, it would be an ideal
potential for getting slow-roll inflation, because of the new
parameter $\epsilon\equiv a_0^4$ which can be made small without any
fine tuning by appropriate choices of the fluxes in (\ref{wf}).
Notice the potential can be approximated as
\beq
	 V \cong 2\epsilon\tau\left(1 - 
	{\epsilon\over r^4}\right) 
\label{Vnice}
\eeq
By simply taking $\epsilon\ll 1$, one can make $V$ as flat as desired.
The $\eta$ slow-roll parameter is 
\beq
		\eta = {V''\over V} \cong -20\epsilon
\eeq
which can easily be made small enough to get 60 e-foldings of
inflation and a nearly scale-invariant spectral index.

\subsection{$\eta$ strikes back}
Unfortunately, the nice potential (\ref{Vnice}) is not the final
answer, because we have ignored the dynamics of the 
overall volume (K\"ahler) modulus $T$.   This is the one 
modulus which is not stabilized by the fluxes.  We will now show that
the interaction of $T$ with the inflaton $\varphi$ induces a large
mass for $\varphi$, which can be expressed as an additional term in the
inflaton potential of the form
\beq
	\delta V = \sfrac m^2\varphi^2,\qquad m^2 \sim V_0 \sim H^2
\eeq
Since $m\sim H$, inflation is spoiled:
\[	\eta \ =\ {V''\over V}  \ \to\ {2\over 3}  \]
The inflaton never rolls slowly!

To understand how the problem arises, we must consider how the
10D metric depends on $T$,
\beq
	ds^2 = e^{-6u}\, a^4\, dx^2+ e^{2u}\, a^{-4}\, 
	\tilde g^{(6)}_{ab} \,	dy^a\, dy^b
\label{newmetric}
\eeq
where $u$ and $T$ are related to the compactification length $L$ 
through
\beq
	e^{4u} = T+\overline T = L^4
\eeq
When (\ref{newmetric}) is used to compute the induced metric that goes into the DBI 
action, the kinetic term of the inflaton gets modified to
\beq	
	(\partial\varphi)^2 \to {(\partial\varphi)^2\over T+\overline
	 T}
\eeq
On the other hand, the low-energy effective action for the brane
position can also be written in the language of  supergravity 
(SUGRA).  Consistency between the DBI and SUGRA approaches implies
that the K\"ahler potential for $T$ gets modified in the presence of 
the D3 brane to 
\beq	
	K = -3\ln(T+\overline T - |\varphi|^2)\equiv - 2\ln 2\sigma
\eeq
In SUGRA, the F-term potential then also gets modified, since
$V \sim e^K$.  This implies that (\ref{Vnice}) is corrected to
\cite{Dewolfe}
\beq
	V\to {V\over (2\sigma)^2} = {V\over (T+\overline T - 
	|\varphi|^2)^2}
\eeq
For small values of $\varphi$ we can expand the new contribution
to obtain the Lagrangian
\beq
	{\cal L} \sim - {(\partial\varphi)^2\over T+\overline T}
	- {V\over (T+\overline T)^2} \left(1 + {2|\varphi|^2\over
	T+\overline T}\right)
\eeq
Because of the new factor $T+\overline T$, we must rescale 
$\varphi$ to get a canonically normalized kinetic term.  Doing so
gives the inflaton mass (in units $M_p=1$) 
\beq
	m^2_\varphi \sim {V\over 2\sigma} \sim H^2
\eeq
Thus the warp factor no longer helps to make $\eta$ small.

\section{Tuning with Superpotential Corrections?}
The solution which was advocated in  ref.\ \cite{KKLMMT} to overcome the
$\eta$ problem was to cancel the unwanted positive contribution to
$m^2_\varphi$ by appropriately modifying the superpotential $W$.  In order
to stabilize the K\"ahler modulus, it was assumed that a 
nonperturbative contribution $Ae^{-aT}$ was present \cite{KKLT},
\beq
		W = W_0 + Ae^{-aT}
\label{kklt}
\eeq
which generates a potential for $T$ with a nontrivial minimum.
Generically one expects this superpotential to also have some
$\varphi$ dependence, which was parametrized in ref.\ \cite{KKLMMT} as
a correction of the form
\beq
	W \to W_0 + Ae^{-aT}(1 + { \delta\varphi^2} )
\eeq
By tuning $\delta$ at the level of  1 part in 100, the inflaton mass
can be made sufficiently small for inflation.   

In an interesting new development, ref.\ \cite{Baumann} noted that it
is not necessary to merely parametrize these corrections; rather, they
can be explicitly computed from string theory.  One can thus check
whether the desired tuning can actually be realized.  To make the
computation tractable, it is necessary to ignore the Calabi-Yau in the
unwarped region and assume that the geometry is well-approximated by
the KS throat by itself.  The superpotential corrections arise due to
the stack of D7 branes wrapping a 4-cycle of the throat, which were a
necessary ingredient of the GKP construction.  This is illustrated
in figure 7.

\begin{figure}[h]
\centerline{
\includegraphics[width=0.4\hsize,angle=0]{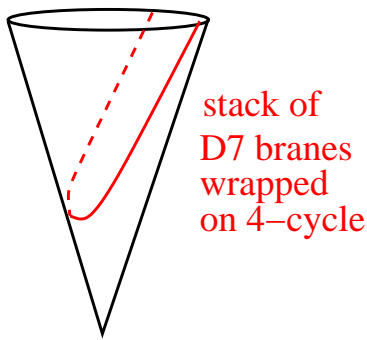}}
\label{fig7}
\caption{D7 branes wrapped on a 4-cycle of the KS throat}
\end{figure}

The superpotential corrections are determined by the 4-cycle
on which the
D7 branes wrap the throat, for which
there are infinitely many choices.  A simple class of 4-cycles which
preserve SUSY is given by \cite{Ouyang}
 \beq
	\prod_{i=1}^4 w_i^{p_i} = \mu^P
\label{ouyang}
\eeq
where $p_i$ are integers, $P=\sum p_i$ and the parameter $\mu$
determines how close to the bottom of the throat the 4-cycle extends.
(Notice that the constraint (\ref{ouyang}), together with the original
conifold restriction (\ref{conifold}), indeed reduces the 8D complex
manifold $\{w_i\}$ to a  4D subspace.) 
Within this class, ref.\ \cite{Baumann} shows that the superpotential
corrections take the form
\beq
	W= W_0+ A e^{-a T}\left(1 - 
	{\prod_{i} w_i^{p_i} \over \mu^P}\right)^{1/N_{D7}}
\label{wcor}
\eeq
where $N_{D7}$ is the number of D7 branes in the stack.

The string-derived correction to $W$ was used in ref.\ \cite{BCDF}
to find the corresponding correction to the F-term potential:
\beqa 
\label{W}
    V_F &=& {\kappa_4^2\over 12\sigma^2}\left[\phantom{\frac13}\!\!\!\!
    (T+\bar T)|W_{,T}|^2\right. -3(\overline W W_{,T} + {\rm c.c.} )
        \\
    &&\qquad\qquad\qquad +\left.\frac32
    \left(\overline W_{,\bar T} w^j W_{,j} + {\rm c.c.}\right)
    +\frac{1}{c} k^{\bar i j} \overline W_{,\bar i} W_{,j}\right] \nonumber \\
    &=& {\kappa_4^2\over 12\sigma^2}\Biggl[ \left[(T+\bar T)a^2+6a\right]
    |A|^2e^{-2a(T+\bar T)} + 3aW_0(Ae^{-aT} + \bar Ae^{-a\bar T})
    \\
    &&\qquad\qquad\qquad
    -\frac32ae^{-a(T+\bar T)}
        \left(\bar A w^j A_{,j} + {\rm c.c.}\right)
        + \frac{1}{c}  k^{\bar \imath j}\overline A_{,\bar \imath} A_{,j}
    e^{-a(T+\bar T)}\Biggr] \,, \nonumber
\label{W2}
\eeqa
The new terms are those in the last line of (\ref{W2}).  This
can be explicitly evaluated in terms of the angular coordinates on
the $T_{1,1}$ manifold, using
\beqa \label{wdef}
    w_1&=&r^{3/2}\, e^{\frac{i}{2} (\psi-\phi_1-\phi_2)} \sin
    \frac{\theta_1}{2} \, \sin \frac{\theta_2}{2} \, ,
    \nonumber\\
    w_2&=&r^{3/2}\,  e^{\frac{i}{2} (\psi+\phi_1+\phi_2)} \cos
    \frac{\theta_1}{2} \, \cos \frac{\theta_2}{2}  \, ,
    \nonumber\\
    w_3&=&r^{3/2}\,  e^{\frac{i}{2} (\psi+\phi_1-\phi_2)} \cos
    \frac{\theta_1}{2} \, \sin \frac{\theta_2}{2}  \, ,
    \nonumber\\
    w_4&=&r^{3/2}\,  e^{\frac{i}{2} (\psi-\phi_1+\phi_2)} \sin
    \frac{\theta_1}{2} \, \cos \frac{\theta_2}{2}  \,,
\eeqa
We find that the new contribution $\delta V_F$ to $V_F$ due to the superpotential
corrections cannot help with tuning the inflaton potential, because
it gets minimized at the value $\delta V_F = 0$ 
when $\theta_1=\theta_2=0$.  For small $\theta_i$,  $\delta V_F$
takes the form
\beq
    \delta V_F = M_{11} (\theta_1^2 + \theta_2^2)
    + M_{12}\cos\left(\sfrac \tilde\psi\right) \theta_1\theta_2 +
    \dots
\eeq
where $\tilde\psi=\psi-\phi_1-\phi_2$ and $M_{11}^2 > \sfracf
M_{12}^2\cos^2(\sfrac\tilde\psi)$ for physically reasonable values of the parameters.
The energetically preferred
brane position is thus at $\theta_i=0$, for which $\delta V_F$ has no
effect.

However, there is another correction which, when combined with
$\delta V_F$, leads to a nonvanishing correction to the potential.
Ouyang \cite{Ouyang} showed that the D7-branes cause the dilaton to
acquire a dependence on position in the compact dimensions:
\beq 
\label{dilaton}
        e^{-\Phi} = {1\over g_s} - {N_{D7}\over 2\pi}
    \log\left({r^{3/2}\over\mu}\sin{\theta_1\over 2}\,
    \sin{\theta_2\over 2}\right) \,.
\eeq
SUSY is not broken by this effect, and so by itself it does not contribute to
the D3 brane potential.  However, if one also introduces nonprimitive
$G_3$ fluxes, which break SUSY spontaneously, (\ref{dilaton}) gets
modified in such a way that the
spatially-varying dilaton background leads to an 
extra contribution to the D3 potential \cite{BCDF, DKS}
\beq
    \delta V_O = - {\delta N(\epsilon)\over 2\pi}\,{T_3 \xi_0^4\over R^2}\,
    \left(r\over r_0\right)^4\,
    \log\left({r^{3/2}\over\mu}\sin{\theta_1\over 2}\sin{\theta_2\over 2}\right) ~ + ~ {\cal O}(\epsilon^2)
\label{ouyang1}
\eeq
Combining this with $\delta V_F$, one sees that $\theta_i= 0$
is no longer a minimum of the full potential since $\delta V_O$
diverges as $\theta_i\to 0$.  There is a competition between the
two terms which leads to nontrivial values of $\theta_i$, at which
the full $\delta V_{\rm tot}$ no longer vanishes.

We can then ask the question: is it possible to tune 
 $\delta V_{\rm tot}$ against the $m^2\varphi^2$ of KKLMMT
to get a flat potential for inflation?  We find that for the class of
embeddings (\ref{ouyang}), the answer is no: the curvature of 
$\delta V_{\rm tot}$ has the wrong sign, and only exacerbates the
$\eta$ problem coming from the $m^2\varphi^2$ term.  Evaluated at
the energetically preferred angles and K\"ahler modulus, 
$\delta V_{\rm tot}$ as a function of $r$ has the form shown in 
figure 8.  It has a maximum at a value of $r=r_{\rm max}$ which is close to the
radius of closest approach of the D7 brane to the bottom of the
throat.  Near this maximum, the curvature of the potential is
negative, but it is much too large to support inflation.  The
contribution to the $\eta$ parameter at this point is
\beq
	\eta \sim {
(T+\overline T) M_p^2\over \tau_3\, r_{\rm max}^2}
	\sim (T+\overline T) g_s (2\pi)^3 \left(M_p\over M_s\right)^4
	\gg 1
\eeq

\begin{figure}
\centerline{\includegraphics[width=0.6\hsize,angle=0]{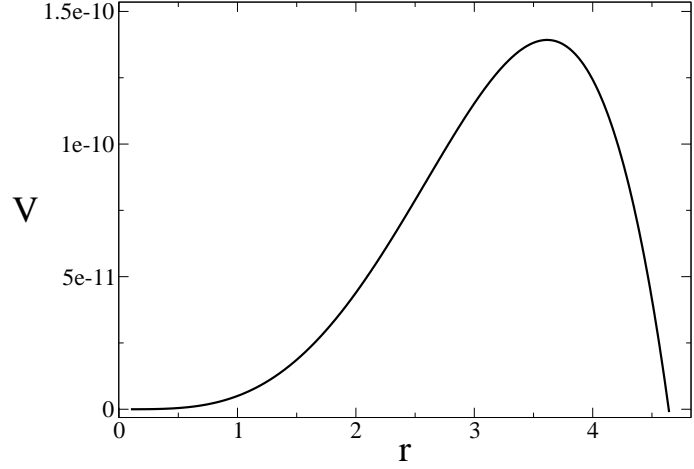}}
\label{fig8}
\caption{The correction to the inflaton
potential due to superpotential
and dilaton corrections, as a function of $r$.}
\end{figure}

Although we do not obtain inflation from this construction, it is
interesting to note that it does give us {\it uplifting}; that is,
$\delta V{\rm tot}$ gives a positive contribution to $V$, which is
necessary for offsetting the negative value of $V_F$ at its minimum,
which comes from the superpotential (\ref{kklt}).  In ref.\ \cite{KKLT}
this problem was overcome by the addition of the \dtb, which
explicitly breaks supersymmetry, and is thus at odds with the SUGRA
formalism used to compute the rest of the potential.  An advantage of
our uplifting contribution is that it does not explicitly break
SUSY, and can thus be derived from a superpotential.

Qualitatively, the uplifting works rather similarly to that from
\dtb branes, as can be seen by comparing the $\sigma$-dependence of
the two potentials: 
\beq
	\delta V_{\overline{{D3}}} = {c\over \sigma^2},\qquad
	\delta V_{\rm tot} = {c\over \sigma^2} \ln(f(\sigma))
\eeq
The shape of the uplifted potential for $\sigma$ is illustrated in
figure 9.

\begin{figure}[h]
\centerline{\includegraphics[width=0.6\hsize,angle=0]{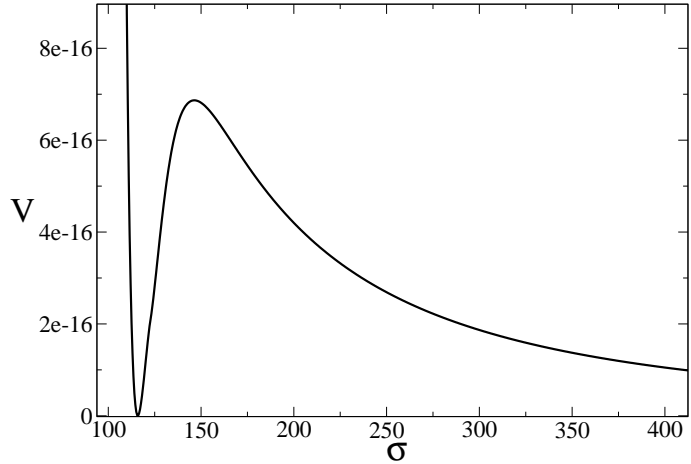}}
\label{fig9}
\caption{Potential for K\"ahler modulus, uplifted to Minkowski vacuum
by superpotential and dilaton corrections.}
\end{figure}

\subsection{Inflation using symmetric throats}
Lest we give the impression that no working models exist, it is
worthwhile to note an exception \cite{IT}, which builds a flat region
into the potential by assuming there are two nearby throats on the
Calabi-Yau.   Clearly a brane at the midpoint between them will be
at an unstable maximum---see figure 10.  The potential has the form
\beq
	V(r) = V_0 -{\tau_3^2 a^8\over \pi^3M_{10}^8}
	\left(|\vec r - \vec r_1|^{-4} + 
	|\vec r + \vec r_1|^{-4} \right)
\eeq
if the two \dtb's at the bottoms of the throats are located at
$\pm \vec r_1$, respectively.  Ref.\ \cite{IT} shows that the
negative curvature of this potential can be tuned against the 
unwanted positive
contribution from the K\"ahler modulus to get $\eta \ll 1$
if 
\beq
	r_1 \sim a_0^{2/3}L
\eeq
which can be naturally achieved.  This therefore looks like a good 
candidate theory for brane-antibrane inflation.

\begin{figure}
\centerline{
\includegraphics[width=0.5\hsize,angle=0]{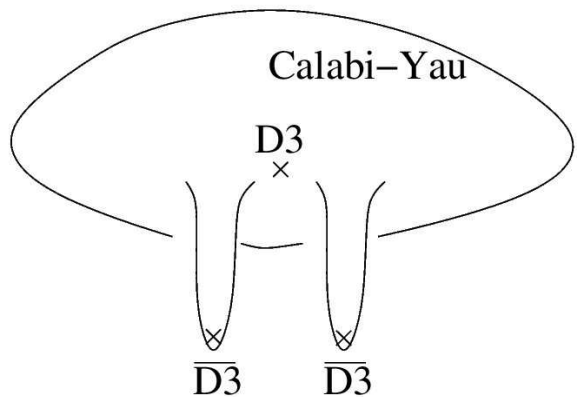}}
\caption{Mobile D3 brane between two throats.}
\label{fig10}
\end{figure}

\section{Conclusions}
Brane-antibrane inflation, which at a qualitative level seems like an
intuitively appealing new way of getting inflation from string theory,
is much harder to successfully implement than one might have guessed.
Even if one is willing to fine-tune the potential, it is not obvious
that string theory provides the latitude to do so, although the
special case of symmetrically-placed throats seems to provide a
working example.  It may also be possible to achieve the desired
tuning by more intricate choices of D7-brane embeddings in the
single-throat scenario \cite{Liam}.


\begin{thebibliography}{99}

\bibitem{JC1}
  J.~M.~Cline,
  ``String cosmology,''
  arXiv:hep-th/0612129.

\bibitem{KKLMMT}
  S.~Kachru, R.~Kallosh, A.~Linde, J.~M.~Maldacena, L.~McAllister and S.~P.~Trivedi,
  ``Towards inflation in string theory,''
  JCAP {\bf 0310}, 013 (2003)
  [arXiv:hep-th/0308055].

\bibitem{DT}
  G.~R.~Dvali and S.~H.~H.~Tye,
  ``Brane inflation,''
  Phys.\ Lett.\  B {\bf 450}, 72 (1999)
  [arXiv:hep-ph/9812483].


\bibitem{BBC}
  N.~Barnaby, C.~P.~Burgess and J.~M.~Cline,
  ``Warped reheating in brane-antibrane inflation,''
  JCAP {\bf 0504}, 007 (2005)
  [arXiv:hep-th/0412040].


\bibitem{cliff}
  C.~P.~Burgess, M.~Majumdar, D.~Nolte, F.~Quevedo, G.~Rajesh and R.~J.~Zhang,
  ``The inflationary brane-antibrane universe,''
  JHEP {\bf 0107}, 047 (2001)
  [arXiv:hep-th/0105204].


\bibitem{GKP}
  S.~B.~Giddings, S.~Kachru and J.~Polchinski,
  ``Hierarchies from fluxes in string compactifications,''
  Phys.\ Rev.\  D {\bf 66}, 106006 (2002)
  [arXiv:hep-th/0105097].

\bibitem{KS}
  I.~R.~Klebanov and M.~J.~Strassler,
  ``Supergravity and a confining gauge theory: Duality cascades and
  $\chi$SB-resolution of naked singularities,''
  JHEP {\bf 0008}, 052 (2000)
  [arXiv:hep-th/0007191].

\bibitem{RS}
  L.~Randall and R.~Sundrum,
  ``A large mass hierarchy from a small extra dimension,''
  Phys.\ Rev.\ Lett.\  {\bf 83}, 3370 (1999)
  [arXiv:hep-ph/9905221].

\bibitem{Dewolfe}
  O.~DeWolfe and S.~B.~Giddings,
  Phys.\ Rev.\  D {\bf 67}, 066008 (2003)
  [arXiv:hep-th/0208123].


\bibitem{KKLT}
  S.~Kachru, R.~Kallosh, A.~Linde and S.~P.~Trivedi,
  ``De Sitter vacua in string theory,''
  Phys.\ Rev.\  D {\bf 68}, 046005 (2003)
  [arXiv:hep-th/0301240].

\bibitem{Baumann}
  D.~Baumann, A.~Dymarsky, I.~R.~Klebanov, J.~Maldacena, L.~McAllister and A.~Murugan,
  ``On D3-brane potentials in compactifications with fluxes and wrapped
  D-branes,''
  JHEP {\bf 0611}, 031 (2006)
  [arXiv:hep-th/0607050].

\bibitem{Ouyang}
  P.~Ouyang,
  ``Holomorphic D7-branes and flavored N = 1 gauge theories,''
  Nucl.\ Phys.\  B {\bf 699}, 207 (2004)
  [arXiv:hep-th/0311084].

\bibitem{BCDF}
  C.~P.~Burgess, J.~M.~Cline, K.~Dasgupta and H.~Firouzjahi,
  ``Uplifting and inflation with D3 branes,''
  JHEP {\bf 0703}, 027 (2007)
  [arXiv:hep-th/0610320].

\bibitem{DKS}
  A.~Dymarsky, I.~R.~Klebanov and N.~Seiberg,
  ``On the moduli space of the cascading SU(M+p) x SU(p) gauge theory,''
  JHEP {\bf 0601}, 155 (2006)
  [arXiv:hep-th/0511254].


\bibitem{IT}
  N.~Iizuka and S.~P.~Trivedi,
  ``An inflationary model in string theory,''
  Phys.\ Rev.\  D {\bf 70}, 043519 (2004)
  [arXiv:hep-th/0403203].

\bibitem{Liam}
  L.\ McAllister, private communication

\end{thebibliography}
\end{document}